\pdfoutput=1
\documentclass[aps,prl,amsmath,floats,floatfix, twocolumn,
superscriptaddress,nofootinbib,showpacs,longbibliography]{revtex4-1}
\usepackage[T1]{fontenc}
\usepackage[utf8]{inputenc}
\usepackage{txfonts}
\usepackage{verbatim}
\usepackage[dvipsnames, usenames]{xcolor}
\definecolor{linkcolor}{rgb}{0.0,0.3,0.5}
\usepackage[hypertexnames=false, unicode, colorlinks=true, linkcolor=linkcolor,
citecolor=linkcolor, filecolor=linkcolor,urlcolor=linkcolor,
pdfusetitle]{hyperref}
\usepackage[all]{hypcap}
\usepackage{graphicx}
\usepackage{xspace}
\usepackage{amssymb}
\usepackage{microtype}

\usepackage{url}

\usepackage[english]{babel}
\usepackage{blindtext}

\usepackage[normalem]{ulem} 
\usepackage{bm} 
\usepackage{mathrsfs}

\usepackage[caption=false]{subfig}

\DeclareMathAlphabet{\mathpzc}{OT1}{pzc}{m}{it}

\usepackage[nomessages]{fp}

\begin{document}
\title{Straightforward mode hierarchy in eccentric binary black hole mergers \\and associated waveform model}
\newcommand{\KITP}{\affiliation{Kavli Institute for Theoretical Physics, University of California Santa Barbara, Kohn Hall, Lagoon Rd, Santa Barbara, California, USA}}
\newcommand{\TAPIR}{\affiliation{Theoretical AstroPhysics Including Relativity and Cosmology, California Institute of Technology, Pasadena, California, USA}}

\author{Tousif Islam}
\email{tislam@kitp.ucsb.edu}
\KITP
\TAPIR

\hypersetup{pdfauthor={Islam et al.}}

\date{\today}

\begin{abstract}
Utilizing publicly available non-spinning eccentric binary black hole (BBH) merger simulations (\href{https://data.black-holes.org/waveforms/catalog.html}{https://data.black-holes.org/waveforms/catalog.html}) from the SXS collaboration~\cite{Hinder:2017sxy}, we present convincing evidence that the waveform phenomenology in eccentric BBH mergers is significantly simpler than previously thought. We find that the eccentric modulations in the amplitudes, phases, and frequencies in different spherical harmonic modes are all related and can be modeled using a single time series modulation. Using this universal eccentric modulation, we provide a model named \texttt{gwNRHME} to seamlessly convert a multi-modal (i.e with several spherical harmonic modes) quasi-circular waveform into multi-modal eccentric waveform if the quadrupolar eccentric waveform is known. This reduces the modelling complexity of eccentric BBH mergers drastically as we now have to model only a single eccentric modulation time-series instead of modelling the effect of eccentricity in all modes. When compared with the NR simulations, our model mismatches are mostly $\sim 10^{-3}$ and are comparable to the numerical errors in the NR simulations. Our method is modular and can be readily added to any quadrupolar non-spinning eccentric waveform model. We make our model publicly available through the \texttt{gwModels} (\href{https://github.com/tousifislam/gwModels}{https://github.com/tousifislam/gwModels}) waveform package.
\end{abstract}

\maketitle

\noindent {\textbf{\textit{Introduction}}.}
While the majority of gravitational wave (GW) signals detected by the LIGO-Virgo-KAGRA collaboration~\cite{Harry_2010,VIRGO:2014yos,KAGRA:2020tym} are with quasi-circular binary black hole (BBH) mergers~\cite{LIGOScientific:2018mvr,LIGOScientific:2020ibl,LIGOScientific:2021usb,LIGOScientific:2021djp}, it is anticipated that some binaries may retain significant eccentricity even as they enter the sensitivity band of current-generation detectors~\cite{Gondan:2020svr,Romero-Shaw:2019itr}. These binaries are likely to originate in dense environments such as globular clusters and galactic nuclei, offering valuable insights into their host environments~\cite{Rodriguez:2017pec,Rodriguez:2018pss,Samsing:2017xmd,Zevin:2018kzq,Zevin:2021rtf}. Claims have been made suggesting that specific events, such as GW190521~\cite{LIGOScientific:2020iuh}, could result from eccentric BBH mergers~\cite{RomeroShaw:2020thy,Gayathri:2020coq,CalderonBustillo:2020xms,Gamba:2021gap}. However, the absence of reliable multi-modal eccentric waveform models poses a substantial challenge in definitively substantiating such findings.
Initial efforts are underway to construct eccentric waveform models by integrating various data-driven and phenomenological techniques, along with insights from post-Newtonian approximations, numerical relativity (NR), and black hole perturbation theory~\cite{Tiwari:2019jtz, Huerta:2014eca, Moore:2016qxz, Damour:2004bz, Konigsdorffer:2006zt, Memmesheimer:2004cv,Hinder:2017sxy, Cho:2021oai,Chattaraj:2022tay,Hinderer:2017jcs,Cao:2017ndf,Chiaramello:2020ehz,Albanesi:2023bgi,Albanesi:2022xge,Riemenschneider:2021ppj,Chiaramello:2020ehz,Ramos-Buades:2021adz,Liu:2023ldr,Huerta:2016rwp,Huerta:2017kez,Joshi:2022ocr,Setyawati:2021gom,Wang:2023ueg,Islam:2021mha,Carullo:2023kvj}. However, providing accurate and fast waveform model for eccentric binaries with various spherical harmonic modes still remains a challenging task. 
Here, utilizing a set of 20 NR simulations of eccentric non-spinning BBH mergers (and additional 3 non-spinning circular BBH merger simulations) (\href{https://data.black-holes.org/waveforms/catalog.html}{https://data.black-holes.org/waveforms/catalog.html}) for binaries with mass ratio $1 \leq q:=m_1/m_2 \leq 3$ (with $m_1$ and $m_2$ being the mass of the larger and smaller black hole, respectively) and eccentricities up to $0.2$ measured $\sim 20$ cycles before the merger~\cite{Hinder:2017sxy}, we demonstrate that the waveform morphology in eccentric BBH mergers is much simpler than previously thought. We note that the modulations due to eccentricity in various quantities such as the amplitudes, phases, and frequencies in different spherical harmonic modes are all related and can be modeled using a single time series modulation. Using this universal eccentric modulation, we present \texttt{gwNRHME} (available through the \texttt{gwModels} package at \href{https://github.com/tousifislam/gwModels}{https://github.com/tousifislam/gwModels}), a model to seamlessly convert a multi-modal quasi-circular spherical harmonic waveform into eccentric waveform spherical harmonic modes if the quadrupolar eccentric waveform is known. Our findings significantly reduce the complexity of modeling eccentric BBH mergers and provide a faster and more accurate framework for building multi-modal NR-faithful waveform models. 

\noindent {\textbf{\textit{Gravitational waveforms}}.}
The gravitational radiation (waveform) from a BBH merger is typically expressed as a superposition of $-2$ spin-weighted spherical harmonic modes:
\begin{align}
h(t, \theta, \phi; \boldsymbol{\lambda}) &= \sum_{\ell=2}^\infty \sum_{m=-\ell}^{\ell} h_{\ell m}(t; \boldsymbol\lambda) ; _{-2}Y_{\ell m}(\theta,\phi),
\label{hmodes}
\end{align}
Here, $(\ell, m)$ are mode indices, and $t$ is the time. The set of parameters describing the binary, such as masses, spins, and eccentricity, is denoted by $\boldsymbol{\lambda}$. The angles ($\theta$,$\phi$) describe the orientation of the binary. Unless otherwise specified, masses (and times) are expressed in geometric units, i.e., we consider $G=c=1$, and all binaries are scaled to have a total mass of $M=1$. Additionally, we restrict ourselves to non-spinning eccentric binaries, so $\boldsymbol{\lambda}:=\{q,e_{\rm ref}\}$ and $e_{\rm ref}$ is the eccentricity estimated at a chosen reference time or frequency.
Each spherical harmonic mode $h^{\ell m}(t; \boldsymbol\lambda)$ is represented as a complex time series and can be further decomposed into a real-valued amplitude $A_{\ell m}(t)$ and phase $\phi_{\ell m}(t)$: $h_{\ell m}(t;q,e) = A_{\ell m}(t) e^{i \phi_{\ell m}(t)}$.
The instantaneous frequency of each spherical harmonic mode is the time derivative of the phase:
\begin{equation}
\omega_{\ell m}(t;q,e_{\rm ref}) = \frac{d\phi_{\ell m}(t)}{dt}.
\label{eq:freq}
\end{equation}
The time coordinate is chosen such that the maximum amplitude of the $(\ell,m)=(2,2)$ mode occurs at $t=0$. Eccentricity of the NR simulations used in this study are measured at a dimensionless frequency of $x_0=(M \omega_{\text{orb,0}})^{2/3}$~\cite{Hinder:2008kv} where $\omega_{\text{orb}}$ is the orbital angular frequency of the binary and can be obtained as: $\omega_{\text{orb}}=0.5 \times \omega_{22}$.

\begin{figure}
\includegraphics[width=\columnwidth]{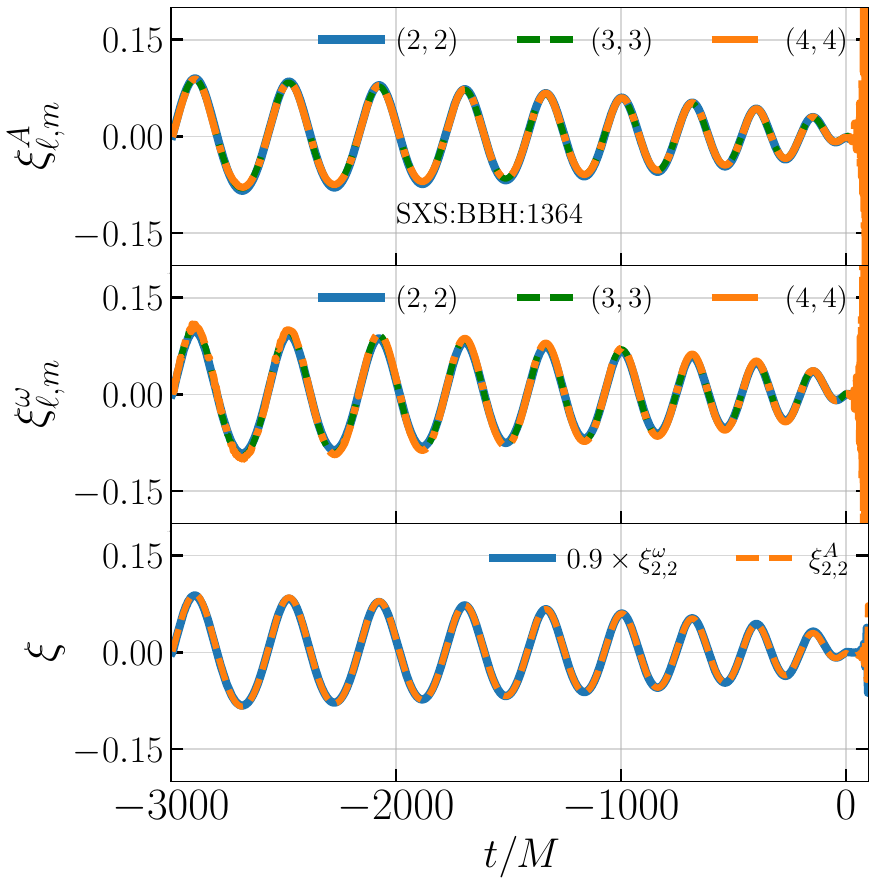}
\caption{We show the eccentric modulations in amplitudes $\xi_{\ell,m}^{A}$ (upper panel; cf. Eq.(\ref{eq:amp_mod})) and in frequencies $\xi_{\ell, m}^{\omega}$ (middle panel; cf. Eq.(\ref{eq:freq_mod})) for three representative modes: $(2,2)$ (blue) $(3,3)$ (green) and $(4,4)$ (orange) for a binary with mass ratio $q=2$ and eccentricity $e_{\text{ref}}=0.05$ measured at a reference dimensionless frequency of $x_{\text{ref}}=0.075$. We extract these modulations from the eccentric NR simulation \texttt{SXS:BBH:1364} and the corresponding circular simulation \texttt{SXS:BBH:0184}~\cite{Hinder:2017sxy}. In the lower panel, we demonstrate that these two modulations are related by a factor of $B=0.9$ (obtained through a phenomenological fit; cf. Eq.(\ref{eq:amp_freq_mod_relation})).}
\label{fig:SXS1364_modulations}
\end{figure}

\begin{figure}
\includegraphics[width=\columnwidth]{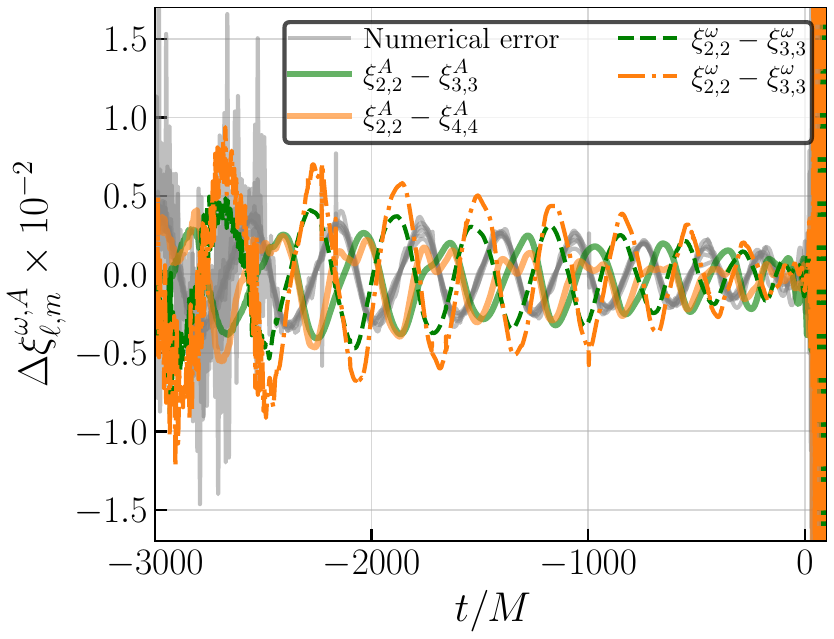}
\caption{We show the differences between eccentric amplitude modulations $\xi_{\ell,m}^{A}$ (solid lines; cf. Eq.(\ref{eq:amp_mod})) and frequency modulations $\xi_{\ell, m}^{\omega}$ (dashed lines; cf. Eq.(\ref{eq:freq_mod})) obtained from different modes (and shown in Figure~\ref{fig:SXS1364_modulations}) for a binary with mass ratio $q=2$ and eccentricity $e_{\text{ref}}=0.05$ measured at a reference dimensionless frequency of $x_{\text{ref}}=0.075$. We extract these modulations from the eccentric NR simulation \texttt{SXS:BBH:1364} and the corresponding circular simulation \texttt{SXS:BBH:0184}~\cite{Hinder:2017sxy}. Additionally, we provide an estimate of numerical errors in both eccentric amplitude modulations and frequency modulations using the highest two resolutions of NR data (grey lines). We find that the differences between the modulations obtained from different modes are comparable to the numerical error in NR.}
\label{fig:SXS1364_modulations_error}
\end{figure}

\noindent {\textbf{\textit{Waveform modulations in eccentric BBH mergers}}.}
Eccentricity introduces additional modulations in the waveforms, affecting various quantities such as amplitude, phase, and frequencies. To quantify these eccentric modulations in a multi-modal eccentric waveform \(h_{\ell m}(t;q,e_{\rm ref})\), we compare it with the corresponding quasi-circular waveform \(h_{\ell m}(t;q,e_{\rm ref}=0)\) in the time domain. These waveforms are initially time-aligned so that \(t=0\) denotes the merger, and they are cast onto the same time grid. Afterward, they are aligned in phase, ensuring that the initial orbital phase is zero. The eccentric modulations are then typically given by so-called `eccentricity estimators' based on either amplitude, phase or frequencies of the $(2,2)$ mode or higher order spherical harmonic modes. The typical form of this eccentricity estimators is~\cite{Mroue:2010re, Healy:2017zqj}:
\begin{equation}\label{eq:23}
e_{X}(t)= b \left(\frac{{X}(t)-{\Bar{X}}(t)}{{\Bar{X}}(t)}\right)
\end{equation}
where $X$ is the amplitude or frequency of the eccentric waveform, $\Bar{X}$ is the corresponding circular feature and $b$ is a constant. The value of $b$ depends on the quantity of interest and the spherical harmonic mode used in extracting the quantity. For example, Ref.~\cite{Healy:2017zqj} finds that $b=\frac{8}{39}$ and $b=\frac{8}{21}$ for $X=A_{22}$ and $X=\omega_{22}$ respectively obtained from the $(2,2)$ mode of the Weyl scalar. Furthermore, replacing the Weyl scalar with quadrupolar mode yields $b=\frac{2}{3}$ for both $X=A_{22}$ and $X=\omega_{22}$~\cite{Healy:2017zqj}. Ref.~\cite{Mroue:2010re} reports that $b=1$ for eccentricity estimators defined based on the proper
separation and coordinate separation between the black holes, and $b=\frac{1}{2}$ when eccentricity estimator is defined based on orbital frequency in the low eccentricity limit. 


\begin{figure*}
\includegraphics[width=\textwidth]{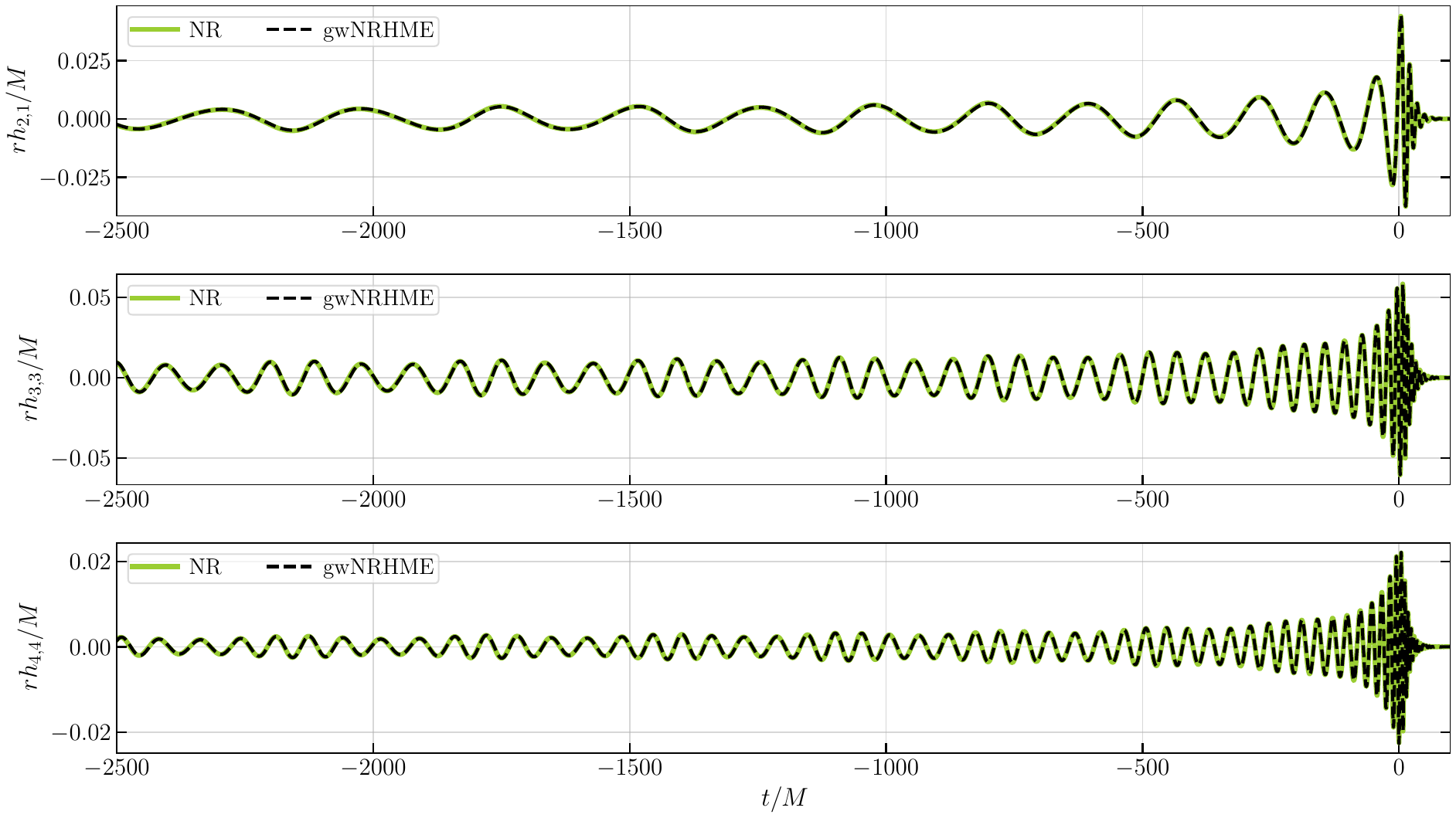}
\caption{We show the eccentric higher-order spherical harmonic modes (black dashed lines) obtained from the \texttt{gwNRHME} model (available at \href{https://github.com/tousifislam/gwModels}{https://github.com/tousifislam/gwModels}) and corresponding NR data from \texttt{SXS:BBH:1371} simulation (green solid lines). We obtain \texttt{gwNRHME} predictions by applying eccentric modulations (computed from the $(2,2)$ mode of
\texttt{SXS:BBH:1371} data; Eq.(\ref{eq:gwNRHME_xi})) on quasi-circular spherical harmonic modes obtained from \texttt{SXS:BBH:0183}. Both simulation is for mass ratio $q=3$. We find that \texttt{gwNRHME} predictions are visually indistinguishable from NR.}
\label{fig:SXS1371_NRHMEcc_wfs}
\end{figure*}

In this paper, we slightly modify the definition of eccentricity estimators.  For the rest of the paper, we drop the term `eccentricity estimator' and use `eccentric modulation' as the latter signifies that these definitions not only is useful to characterize eccentricity but also to model the extra oscillatory features observed in eccentric BBH waveforms. We define the eccentric frequency modulation for each mode as:
\begin{equation}
\xi_{\ell m}^{\omega}(t;q,e_{\rm ref}) = b_{\ell m}^\omega \frac{\omega_{\ell m}(t;q,e_{\rm ref})-\omega_{\ell m}(t;q,e_{\rm ref}=0)}{\omega_{\ell m}(t;q,e_{\rm ref}=0)}.
\label{eq:freq_mod}
\end{equation}
For the amplitude, we define the modulations as:
\begin{equation}
\xi_{\ell m}^{A}(t;q,e_{\rm ref}) = b^{A}_{\ell m} \frac{2}{\ell} \frac{A_{\ell m}(t;q,e_{\rm ref})-A_{\ell m}(t;q,e_{\rm ref}=0)}{A_{\ell m}(t;q,e_{\rm ref}=0)}.
\label{eq:amp_mod}
\end{equation}
Note that while $\xi_{\ell m}^{\omega}(t;q,e_{\rm ref})$ does not have any mode dependence, $\xi_{\ell m}^{A}(t;q,e_{\rm ref})$ depends on the $\ell$ value of the spherical harmonic mode. We set the constants $b_{\ell m}^\omega$ and $b^{A}_{\ell m}$ to be unity for now.

\noindent {\textbf{\textit{Mode hierarchy in eccentric BBH mergers}}.}
We discover that the mode hierarchy in eccentric BBH mergers is much simpler than previously thought. In fact, the modulations induced by eccentricity in a single mode for any quantity (such as amplitude, phase, or frequency) can be used to provide the modulations in all quantities in all spherical harmonic modes. We demonstrate this for an eccentric BBH simulation \texttt{SXS:BBH:1364}~\cite{Hinder:2017sxy} characterized by a mass ratio $q=2$ and eccentricity $e_{\text{ref}}=0.05$ measured at a reference dimensionless frequency of $x_{\text{ref}}=0.075$. We show that the amplitude modulations obtained from different spherical harmonic modes are the same (Figure~\ref{fig:SXS1364_modulations}; upper panel). Additionally, the frequency modulations in different spherical harmonic modes are also the same (Figure~\ref{fig:SXS1364_modulations}; middle panel). Furthermore, the amplitude modulations and frequency modulations are related by a scaling factor $B$ (Figure~\ref{fig:SXS1364_modulations}; lower panel) such that
\begin{equation}
\label{eq:amp_freq_mod_relation}
\xi_{\ell m}^{A}(t;q,e_{\rm ref}) = B \xi_{\ell m}^{\omega}(t;q,e_{\rm ref}).
\end{equation}
We obtain $B=0.9$ for all non-spinning eccentric binaries considered in this paper.
This means that the constants $b_{\ell m}^{A}$ and $b_{\ell m}^\omega$ are not equal (as we have initially guessed in Eq.(\ref{eq:freq_mod}) and Eq.(\ref{eq:amp_mod})). This difference shows up in $B$.

\begin{figure*}
\includegraphics[width=\textwidth]{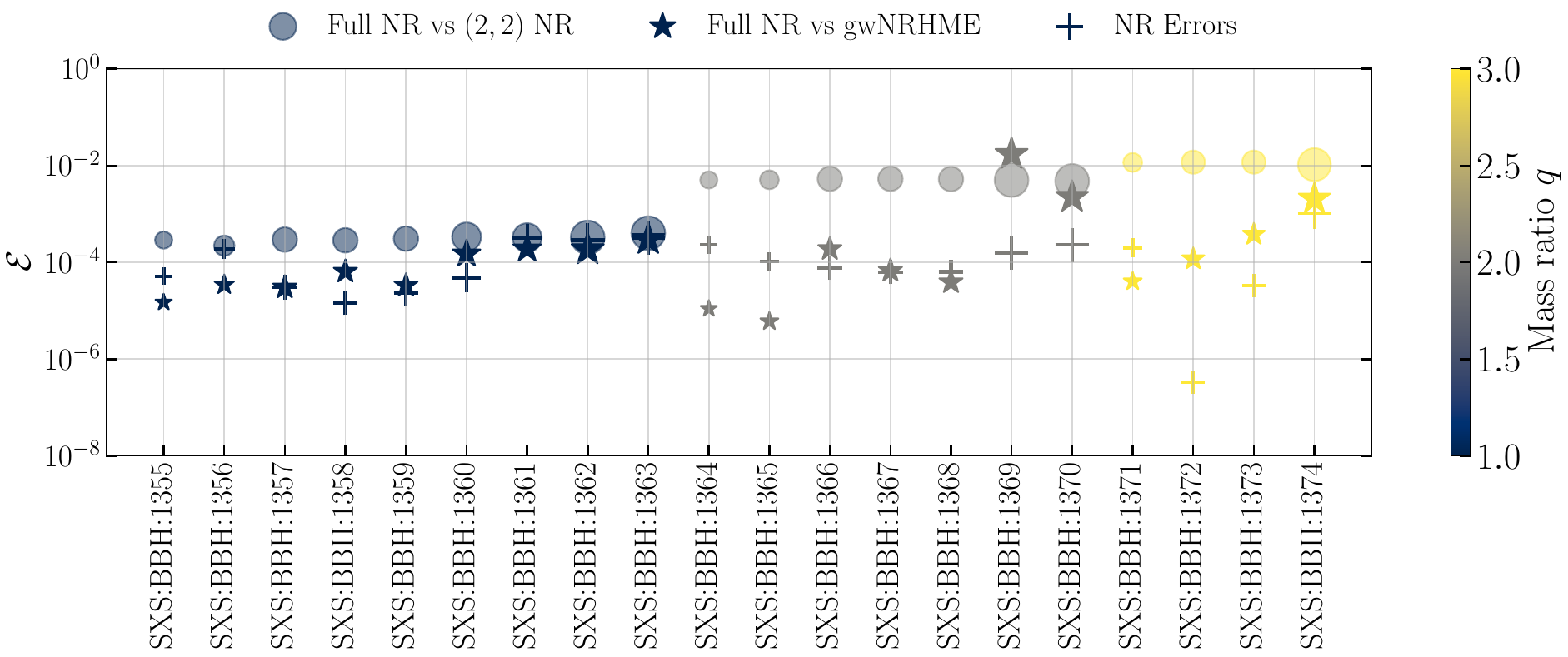}
\caption{We show the relative $L_2$-norm error (defined in Eq.(\ref{eq:l2err})) between the full NR eccentric waveform and the $(\ell,m)=(2,2)$ mode data as blue circles. Additionally, we present the relative $L_2$-norm error between the full NR eccentric waveform and the \texttt{gwNRHME} model (available at \href{https://github.com/tousifislam/gwModels}{https://github.com/tousifislam/gwModels}) prediction as stars. We also provide numerical errors in NR simulations obtained by comparing full NR data between simulations with the highest two resolutions ("+" marker). Markers are color-coded according to the mass ratio values of the simulations, and marker size increases with eccentricity.}
\label{fig:Modelling_errors}
\end{figure*}

We have further confirmed that the differences between $\xi_{\ell m}^{A}(t;q,e_{\rm ref})$ (and $\xi_{\ell m}^{\omega}(t;q,e_{\rm ref})$) computed from different spherical harmonic modes are comparable to the numerical errors in those NR simulations. We obtain the numerical errors by computing the differences between the respective eccentric modulations obtained from the highest two resolution of the NR simulations. Figure~\ref{fig:SXS1364_modulations_error} demonstrates this for \texttt{SXS:BBH:1364} where the differences between $\xi_{\ell m}^{A}(t;q,e_{\rm ref})$ (and $\xi_{\ell m}^{\omega}(t;q,e_{\rm ref})$) computed from different spherical harmonic modes closely follow the NR error.

\noindent {\textbf{\textit{Construction of the  \texttt{gwNRHMEcc} waveform model}}.}
Our results greatly simplify the modeling challenges in eccentric binaries. They indicate that we can now develop a multi-modal waveform model for eccentric binaries if we already have a multi-modal waveform model for non-spinning quasi-circular BBH mergers and an eccentric waveform model for the dominant quadrupolar mode of the waveform. To achieve this, we now present \texttt{gwNRHME}, a model that takes an eccentric $(2,2)$ mode time-domain waveform $h^{\tt{model}}(t; q, e_{\rm ref})$ and the corresponding multi-modal quasi-circular waveform $h^{\tt{model}}_{\ell,m}(t; q, e_{\rm ref}=0)$ as input. 
We immediately compute the amplitude and frequency [$A^{\tt{model}}(t; q, e_{\rm ref})$, $\omega^{\tt{model}}(t; q, e_{\rm ref})$] of the eccentric $(2,2)$ mode waveform. Similarly, we obtain the amplitudes and frequencies [$A^{\tt{model}}_{\ell,m}(t; q, e_{\rm ref}=0)$,$\omega^{\tt{model}}_{\ell,m}(t; q, e_{\rm ref}=0)$] of different spherical harmonic modes in quasi-circular waveform.

\begin{figure}
\includegraphics[width=\columnwidth]{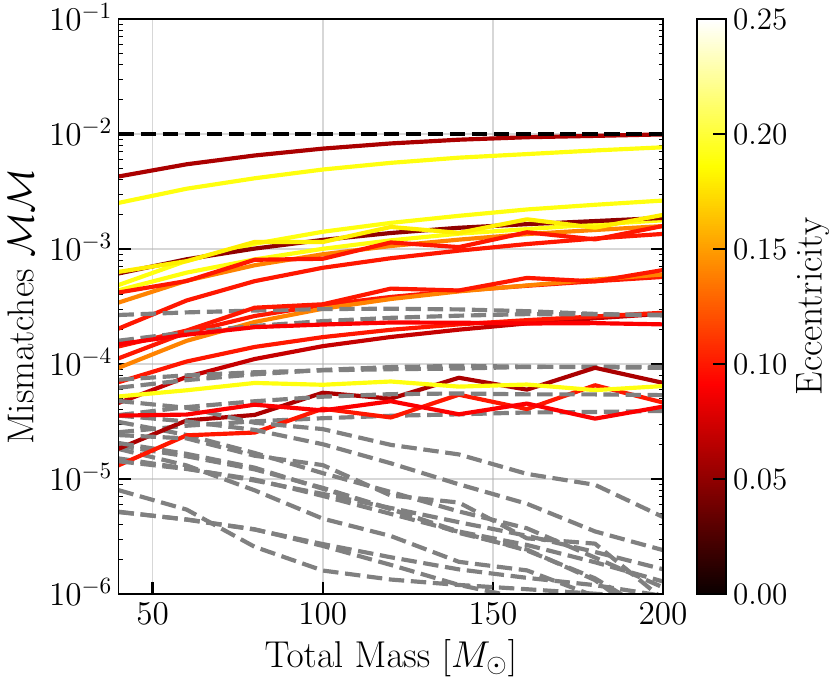}
\caption{We show the frequency-domain average mismatches (defined in Eq.(\ref{Eq:freq_domain_Mismatch})) between full NR eccentric waveform and the \texttt{gwNRHME} model (available at \href{https://github.com/tousifislam/gwModels}{https://github.com/tousifislam/gwModels}) prediction for all 20 eccentric BBH simulations considered in this paper. The mismatches are shown as a function of the detector-frame total mass of the binary $M$ at orbital phase $\phi=0.0$, and are computed using the advanced LIGO design sensitivity noise curve. We set the minimum (maximum) frequency to be 20Hz (990Hz). The dashed horizontal line demarcates a mismatch of 0.01, a commonly used threshold for sufficiently good model quality. Dashed grey lines represents the mismatches between NR simulations with highest two resolutions. Brighter yellow color represents larger eccentricity and brighter red denote smaller values of eccentricity.}
\label{fig:mismatches}
\end{figure}

Input eccentric $(2,2)$ mode serves as the final $(2,2)$ mode with no additional processing:
\begin{equation}
\label{eq:gwNRHME_xi}
h_{22}^{\tt{gwNRHME}}(t; q, e_{\rm ref}) = h^{\tt{model}}(t; q, e_{\rm ref}).
\end{equation}
We then compute the eccentric amplitude modulation $\xi_{\ell m}^{A}(t; q, e_{\rm ref})$ from the $(2,2)$ mode as:
\begin{equation}
\xi_{\ell m}^{A}(t; q, e_{\rm ref}) = \frac{A^{\tt{model}}(t;q,e_{\rm ref})-A^{\tt{model}}(t;q,e_{\rm ref}=0)}{A^{\tt{model}}(t; q, e_{\rm ref}=0)}
\end{equation}
and call it $\xi(t) := \xi_{22}^{A}(t; q, e_{\rm ref})$.
Exploiting the relations found earlier in Eq.(\ref{eq:freq_mod}), Eq.(\ref{eq:amp_mod}) and Eq.(\ref{eq:amp_freq_mod_relation}), we compute the amplitude of higher order modes as:
\begin{equation}
A_{\ell m}^{\tt{gwNRHME}}(t; q, e_{\rm ref}) = A_{\ell m}(t; q, e_{\rm ref}=0) \Big[ \frac{\ell}{2} \xi(t) + 1 \Big].
\end{equation}
The frequency of different spherical harmonic modes is then simply:
\begin{equation}
\omega_{\ell m}^{\tt{gwNRHME}}(t; q, e_{\rm ref}) = \omega_{\ell m}(t; q, e_{\rm ref}=0) \Big[ \frac{\xi(t)}{B} + 1 \Big].
\end{equation}
We then integrate the frequency to get the phase of each mode:
\begin{equation}
\phi_{\ell m}^{\tt{gwNRHME}}(t; q,  e_{\rm ref}) = \phi_{0} + \int \omega_{\ell m}^{\tt{gwNRHME}}(t; q, e_{\rm ref}) dt.
\end{equation}
Here, $\phi_{0}$ is the integration constant, and its value is set to be the initial value of the respective quasi-circular phase $\phi_{\ell m}(t; q, _{\rm ref}=0)$. We then combine the amplitude and phase terms to obtain complex time-series for each mode as:
\begin{equation}
h_{\ell m}^{\tt{gwNRHME}}(t; q, e_{\rm ref}) = A_{\ell m}^{\tt{gwNRHME}}(t; q, e_{\rm ref}) e^{\phi_{\ell m}^{\tt{gwNRHME}}(t;q, e_{\rm ref})}.
\end{equation}
Our model can be readily used to extend current $(2,2)$ mode waveform models for eccentric non-spinning BBH mergers and incorporate higher-order spherical harmonic modes.

\noindent {\textbf{\textit{Modelling accuracy of \texttt{gwNRHME} waveform model}}.}
To demonstrate the effectiveness of our waveform model \texttt{gwNRHME}, we utilize the $(2,2)$ mode of all eccentric SXS NR simulations presented in Ref.~\cite{Hinder:2017sxy} and their corresponding quasi-circular simulations. We then convert all higher-order spherical harmonic modes in quasi-circular simulations into eccentric spherical harmonic modes. Finally, we compare these converted eccentric spherical harmonic modes to actual SXS NR data. We find that \texttt{gwNRHME} can predict the higher order spherical harmonic modes very accurately. We demonstrate this for \texttt{SXS:BBH:1371} in Figure~\ref{fig:SXS1371_NRHMEcc_wfs}. We utilize the $(2,2)$ mode eccentric spherical harmonic mode extracted from \texttt{SXS:BBH:1371} and all quasi-circular spherical harmonic modes from \texttt{SXS:BBH:0183} to predict the eccentric higher order spherical harmonics. We find that \texttt{gwNRHME} predictions are visually indistinguishable from NR.

To assess the modeling accuracy of \texttt{gwNRHHE}, we calculate a time/phase optimized time-domain relative $L_2$-norm between NR data and \texttt{gwNRHHE} predictions, as well as the noise-curve-weighted frequency-domain mismatches between them. The relative $L_2$-norm between two waveforms $h_1(t)$ and $h_2(t)$ is given by:
\begin{equation}\label{eq:l2err}
\mathcal{E} = \int_{t_{\rm min}}^{t_{\rm max}} \frac{|h_{1}(t) - h_{2}(t)|^2}{|h_{1}(t)|^2} dt,
\end{equation}
where $t_{\rm min}$ and $t_{\rm max}$ represent the initial and final times of the waveforms. The frequency domain mismatch between two waveforms $h_1$ and $h_2$ is defined as:
\begin{gather}
	\mathcal{M} \mathcal{M} = 1 - 4 \mathrm{Re}
	\int_{f_{\mathrm{min}}}^{f_{\mathrm{max}}}
	\frac{\tilde{h}_{1} (f) \tilde{h}_{2}^* (f) }{S_n (f)} df,
	\label{Eq:freq_domain_Mismatch}
\end{gather}
where $\tilde{h}(f)$ indicates the Fourier transform of the complex strain $h(t)$, and $^*$ indicates complex conjugation. We choose $S_n(f)$ to be the design sensitivity curve of the advanced LIGO detector~\cite{Harry_2010}. We set $f_{\rm min}=20Hz$ and $f_{\rm max}=999Hz$. In cases where the initial frequency of the time-domain NR data is larger than $20$Hz, we set $f_{\rm min}$ to be the starting frequency of the NR data. For each binary, we compute the mismatches for five different inclinations ($\iota=[0,\pi/6,\pi/4,\pi/3,\pi/2]$) while keeping the orbital phase value to be zero and report the average.

\begin{figure}
\includegraphics[width=\columnwidth]{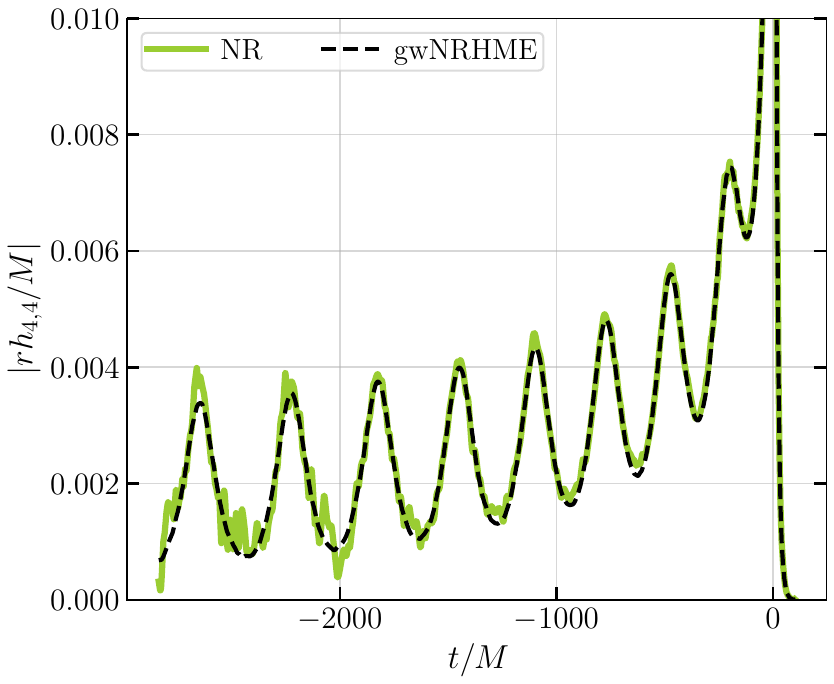}
\caption{We demonstrate that \texttt{gwNRHME} model (black dashed line; available at \href{https://github.com/tousifislam/gwModels}{https://github.com/tousifislam/gwModels}) can be used to provide clean eccentric higher-order spherical harmonic modes. We illustrate this for \texttt{SXS:BBH:1374} (green), characterized by a mass ratio $q=3$ and eccentricity $e_{\rm ref}=0.19$ at a reference dimensionless frequency of $x_{\text{ref}}=0.075$. While the amplitude of the $(\ell,m)=(4,4)$ mode obtained from NR is quite noisy, the \texttt{gwNRHME} prediction (obtained by combining $(2,2)$ mode eccentric modulation from \texttt{SXS:BBH:1374} simulation with quasi-circular \texttt{SXS:BBH:0183} simulation) is clean and goes right through the NR data.}
\label{fig:smooth_NRHME}
\end{figure}

In Figure~\ref{fig:Modelling_errors}, we show the relative $L_2$-norm errors between (i) the full NR eccentric waveform~\footnote{For the full waveform, we consider $(2,1)$,$(2,2)$,$(3,2)$,$(3,3)$,$(4,3)$,$(4,4)$ modes.} and the $(\ell,m)=(2,2)$ mode data (circles), (ii) the full NR eccentric waveform and the \texttt{gwNRHME} predictions for different modes (stars), and (iii) the full NR eccentric waveforms obtained from the highest two resolution NR simulations ("+" marker). The latter provides an estimate of the numerical error in NR simulations. We find that obtaining higher-order eccentric harmonics modes using \texttt{gwNRHME} greatly improves the relative $L_2$-norm errors by two to three orders of magnitude. In most of the binaries, once higher-order modes are added using \texttt{gwNRHME}, the relative $L_2$-norm between actual NR data and \texttt{gwNRHME} predictions becomes comparable to the numerical errors in NR itself. Next, we compute the average frequency-domain mismatches between the full NR eccentric waveform and the \texttt{gwNRHME} prediction for all 20 eccentric BBH simulations. We find that the mismatches are always below $0.01$, a commonly used threshold to indicate good model quality (Figure~\ref{fig:mismatches}). We note that for most of the binaries, mismatch values are comparable to the mismatches between the two highest resolution NR simulations.

We also note that \texttt{gwNRHME} can be used to provide clean eccentric higher-order spherical harmonic modes in NR simulations. We illustrate this for \texttt{SXS:BBH:1374} (Figure~\ref{fig:smooth_NRHME}), characterized by a mass ratio $q=3$ and eccentricity $e_{\rm ref}=0.19$ at a reference dimensionless frequency of $x_{\text{ref}}=0.075$~\cite{Hinder:2017sxy}. Due to the large mass ratio and eccentricity values, the extracted higher-order spherical harmonic modes (e.g. $(4,4)$ mode shown in Figure~\ref{fig:smooth_NRHME}) are noisy. However, when we combine the eccentric modulation in the $(2,2)$ mode of \texttt{SXS:BBH:1374} with the quasi-circular \texttt{SXS:BBH:0183} simulation, the resultant \texttt{gwNRHME} prediction for the $(4,4)$ mode exhibits noise-free amplitude and accurately tracks the noisy NR data. Our model can therefore be used as an alternative to filter out noise in higher order spherical harmonics modes extracted from current eccentric BBH NR simulations.

\noindent {\textbf{\textit{Discussion \& Conclusion}}.}
In this paper, we provide simple relations between the quadrupolar mode of the gravitational waveform in eccentric BBH mergers and higher order spherical harmonic modes. In particular, we show that the eccentricity-induced modulations in amplitudes and frequencies in different spherical harmonic modes are all related by a constant factor. This allows us to build a waveform model named \texttt{gwNRHME} that can exploit the $(2,2)$ mode eccentric waveform along with its corresponding multi-modal quasi-circular waveforms to provide multi-modal eccentric waveforms. Our model is publicly available through \texttt{gwModels} waveform package (\href{https://github.com/tousifislam/gwModels}{https://github.com/tousifislam/gwModels}) and can be immediately used to extend current quadrupolar mode non-spinning eccentric waveform models such as \texttt{EccentricTD}~\cite{Tanay:2016zog} or \texttt{EccentricIMR}~\cite{Hinder:2017sxy,Cho:2021oai}. Our method will significantly reduce the modelling complexity of eccentric BBH mergers and help in developing reliable multi-modal eccentric models using different formalism including NR and EOB within a short span of time. Another interesting direction is to investigate similar scalings for the frequency domain eccentric waveforms and extend the framework to binaries with spins. We leave these for immediate future.

\noindent {\textbf{\textit{Acknowledgments}}.}
We thank Scott Field, Gaurav Khanna, Chandra Kant Mishra, Harald P. Pfeiffer and Tejaswi Venumadhav for fruitful discussions and comments on the manuscript. We are thankful to Ian Hinder, Lawrence E. Kidder, and Harald P. Pfeiffer for making the NR simulations~\cite{Hinder:2017sxy} publicly available, and to the SXS collaboration for maintaining publicly available catalog of NR simulations. This research was supported in part by the National Science Foundation under Grant No. NSF PHY-2309135 and the Simons Foundation (216179, LB).

\bibliography{References}

\end{document}